\documentclass[12pt]{article}

\usepackage{epsf,amsfonts,hyperref}
\renewcommand{\appendix}[1]{
    \addtocounter{section}{1}
    \setcounter{equation}{0}
    \renewcommand{\thesection}{\Alph{section}}
    \section*{Appendix \thesection\protect\indent #1}
    \addcontentsline{toc}{section}{Appendix \thesection\ \ \ #1}
}

\newcommand\lto {\leadsto}
\newcommand\1 {{\mathbf 1}}

\renewcommand\l{\lambda}
\renewcommand\L{\Lambda}

\newcommand{\Mdet}{{\,\rm Mdet}}
\newcommand{\Tr}{{\,\rm Tr}\:}

\newcommand{\diag}{{\,\rm diag}}
\newcommand{\ovl}{\overline}

\newcommand{\Sym}{{\Sigma}}

\newcommand{\td}[1]{{\tilde{#1}}}

\newcommand{\om}{\omega}

\newcommand{\ee}[1]{{{\rm e}^{#1}}}

\renewcommand{\d}{{{\partial}}}

\newcommand{\Pint}{{\int\kern -1.em -\kern-.25em}}

\renewcommand{\Re}{{\mathrm{Re}}}
\renewcommand{\Im}{{\mathrm{Im}}}

\renewcommand{\and}{{\qquad {\rm and} \qquad}}

\newcommand{\virg}{{\,\, , \qquad}}

\newcommand{\beq}{\begin{equation}}
\newcommand{\eeq}{\end{equation}}
\newcommand{\bea}{\begin{eqnarray}}
\newcommand{\eea}{\end{eqnarray}}
\newcommand{\bacc}{\left\{ \begin{array}{l}}
\newcommand{\eacc}{\end{array}\right.}
\renewcommand{\thesection}{\arabic{section}}

\makeatletter
\@addtoreset{equation}{section}
\makeatother
\newtheorem{theorem}{Theorem}[section]

\newtheorem{remark}{Remark}[section]
\newtheorem{proposition}{Proposition}[section]
\newtheorem{lemma}{Lemma}[section]
\newtheorem{corollary}{Corollary}[section]
\newtheorem{definition}{Definition}[section]

\def\br{\begin{remark}\rm\small}
\def\er{\end{remark}}
\def\bt{\begin{theorem}}
\def\et{\end{theorem}}
\def\bd{\begin{definition}}
\def\ed{\end{definition}}
\def\bp{\begin{proposition}}
\def\ep{\end{proposition}}
\def\bl{\begin{lemma}}
\def\el{\end{lemma}}
\def\bc{\begin{corollary}}
\def\ec{\end{corollary}}
\def\beaq{\begin{eqnarray}}
\def\eeaq{\end{eqnarray}}
\newcommand{\proof}[1]{{\noindent \bf proof:}\par
{#1} $\square$}

\newcommand{\R}{{\bf\rm  R}}
\newcommand{\C}{{\bf\rm  C}}

\newcommand\encadremath[1]{\vbox{\hrule\hbox{\vrule\kern8pt
\vbox{\kern8pt \hbox{$\displaystyle #1$}\kern8pt}
\kern8pt\vrule}\hrule}}
\def\enca#1{\vbox{\hrule\hbox{
\vrule\kern8pt\vbox{\kern8pt \hbox{$\displaystyle #1$}
\kern8pt} \kern8pt\vrule}\hrule}}

\begin{document}

\sloppy


\pagestyle{empty}
\hfill SPT-05/011, UB-ECM-PF 05/03
\addtolength{\baselineskip}{0.20\baselineskip}
\vspace{56pt}
\begin{center}
{\Large
2-matrix versus complex matrix model,

integrals over the unitary group as triangular integrals
}

\vspace{10pt}
{\sl B.\ Eynard}\hspace*{0.05cm}\footnote{ E-mail: bertrand.eynard@cea.fr }\\
\vspace{6pt}
Service de Physique Th\'{e}orique de Saclay,
 CEA/DSM/SPhT - CNRS/SPM/URA 2306,\\
F-91191 Gif-sur-Yvette Cedex, France.\\

{\sl A.\ Prats Ferrer}\hspace*{0.05cm}\footnote{ E-mail: prats@ecm.ub.es }\\
\vspace{6pt}
Universitad Barcelona, Departament d'Estructura i constituents de la Mat\`eria.\\
Av. Diagonal 647, 08028, Barcelona.\\

\vspace{20pt}
{\bf Abstract}

We prove that the 2-hermitean matrix model and the complex-matrix model obey the same loop equations,
and as a byproduct, we find a formula for Itzykzon-Zuber's type integrals over the unitary group.
Integrals over $U(n)$ are rewritten as gaussian integrals over triangular matrices and then computed explicitely.
That formula is an efficient alternative to the former Shatashvili's formula.
\end{center}
%





\newpage
\pagestyle{plain}
\setcounter{page}{1}

\section{Introduction}

It has been noticed for a long time now, that the so called ''Two-Hermitean-Matrix-Model'' (introduced in particular for quantum gravity \cite{kazakov, ZJDFG}) and the
so called ''Complex-Matrix-Model'' (used in particular for its applications to Laplacian growth models \cite{WiegZab, Zab}, and string theory \cite{BMN}) share lots of similarities:
They have the same leading large $N$ expansion properties,
and, both are associated to some ensembles of biorthogonal polynomials which have formaly the same properties.
Here, we add a new piece to make this correspondence more precise, we prove that both models have the same loop equations.

Both models are not defined for the same weights, in fact, the set of weights for which one model is well defined
has no intersection with the set of weights for which the other model is well defined.
However, each model can be analyticaly continued to a larger set of weights, and in that sense, the two models coincide.

When written in terms of eigenvalues, this identification of the 2-hermitean-matrix-model and complex-matrix-model has some interesting corolary:
it gives a formula for computing integrals (of the Itzykzon-Zuber type) over the unitary group, as gaussian integrals over triangular matrices.
Therefore, we obtain a very explicit formula for all correlators of the Shatashvili's type \cite{shata}.
In \cite{shata} S. Shatashvili found a formula for all $U(n)$ correlation functions, but his formula still contains integrals, is not explicitely symmetric in all variables, and is very difficult to use
for practical purposes, such as \cite{BEmixed}. In the particular case of the 2-point correlation function, Morozov has found a much simpler formula \cite{morozov}.
In \cite{morozov} A. Morozov computed it for $U(n)$ with $n\leq 3$ and conjectured it for $n>3$. Morozov's formula was later proven for all $n$ in \cite{BEmixed}, and
written in an even simpler form \cite{eynmorozov}.
Here, we find a natural generalization of Morozov's formula.
The formula we find here, contains no integration, it gives the $U(n)$ correlation functions as the sum of a finite number of terms,
and is very efficient for effective computations.
It also provides an alternative new proof of Itzykzon-Zuber's formula.

The derivations proposed in this article are elementary, and it would be interesting to put them in the more general framework of group representation theory \cite{Knapp, fulton}.

\bigskip

The main results presented in this paper are:

$\bullet$ Theorem \ref{theoremHC} and in particular Remark \ref{remidZHZC}, which states the equivalence between the Hermitean-2-matrix model and the complex-matrix model:
\beq
\int_{H_n\times H_n} dM_1\, dM_2\, F(M_1,M_2)\,\ee{-\gamma \Tr M_1 M_2}
\equiv
\int_{GL_n(\C)} dZ\, F(Z,Z^\dagger)\,\ee{-\gamma \Tr Z Z^\dagger}
\eeq
The definitions of each terms and the meaning of that equality are explained in section \ref{secrelZHZC}.

$\bullet$ Theorem \ref{ThmintUintT}, which allows to compute $U(n)$ integrals as triangular integrals.
\bea
&&  \int_{U(n)} dU\, F(X,U Y U^\dagger)\, \ee{- \Tr X U Y U^\dagger} \cr
&\propto& {\sum_\sigma \sum_\tau (-1)^\sigma (-1)^\tau \ee{- \Tr X_\sigma Y_\tau}  \int_{T(n)} F(X_\sigma+T,Y_\tau+T^\dagger)\, \ee{- \Tr T T^\dagger}
\over \Delta(X)\Delta(Y) }
\eea
for any polynomial invariant function $F$.

$\bullet$ Theorem \ref{ThmWPiM}, which gives a formula for computing triangular matrix gaussian integrals.
We parametrize polynomial invariant functions by pairs of permutations (of some size $R$), and a basis is written $F_{\pi,\pi'}$.
Theorem \ref{ThmWPiM} gives the result of integration over triangular matrices:
\bea\label{introintFpiM}
&& {\int_{T(n)} dT\,\ee{-\Tr T T^\dagger}\, F_{\pi,\pi'}(\vec{x},\vec{y},X+T,Y+T^\dagger)
\over \int_{T(n)} dT\, \ee{-\Tr T T^\dagger}} \cr
&=& \left({\cal M}^{(R)}(\vec{x},\vec{y},X_n,Y_n)\, {\cal M}^{(R)}(\vec{x},\vec{y},X_{n-1},Y_{n-1})\, \dots  {\cal M}^{(R)}(\vec{x},\vec{y},X_1,Y_1)       \right)_{\pi,\pi'}
\eea
where ${\cal M}^{(R)}(\vec{x},\vec{y},X_n,Y_n)$ is the matrix of size $R!$, indexed by pairs of permutations:
\beq
{\cal M}^{(R)}_{\pi,\rho}(\vec{x},\vec{y},X_n,Y_n) = \prod_{i=1}^R \left(\delta_{\pi(i),\rho(i)}+{1\over (x_i-X_n)(y_{\pi(i)}-Y_n)}\right)
\eeq
Theorem \ref{thmMcommute} shows that the matrices in eq.\ref{introintFpiM} commute together, and can be simultaneousy diagonalized.

$\bullet$ Theorem \ref{thmmixedcor}, which gives a formula for computing correlation functions in terms of biorthogonal polynomials:
\bea
 {\int_{H_n\times H_n}\, dM_1\, dM_2\, F_{\pi,\pi'}(\vec{x},\vec{y},M_1,M_2)\, \ee{-\Tr(V_1(M_1)+V_2(M_2)+ M_1 M_2)}
\over \int_{H_n\times H_n}\, dM_1\, dM_2\, \ee{-\Tr(V_1(M_1)+V_2(M_2)+\gamma M_1 M_2)}}
= \left( \Mdet \left({\cal M}^{(R)}(\vec{x},\vec{y},Q,P^t)\right)\right)_{\pi,\pi'} \cr
\eea
where notations are explained in section \ref{mixedcor}.

\bigskip

\underline{Outline:}

$\bullet$ In part 2 we give definitions of groups and measures.

$\bullet$ In part 3, we prove the equivalence between the Hermitean-2-matrix model and the complex-matrix model, by showing that they have the same loop equations.

$\bullet$ In part 4, we prove the identity between $U(n)$ integrals and triangular integrals, and give some examples.
In particular we rederive Itzykson-Zuber's formula and Morozov's formula.

$\bullet$ In part 5, we compute the triangular integrals, by parametrizing polynomial invariant functions with pairs of permutations.
In particular we explicit all four point functions.

$\bullet$ In part 6, we integrate over eigenvalues using biorthogonal polynomials technics, and get expressions for correlation functions.

\bigskip

\section{Definitions}

\subsection{Ensembles}

Let

$\bullet$ $U(n):=$ group of $n\times n$ unitary matrices, with the normalized Haar measure.

$\bullet$ $H_n:=$ group of $n\times n$ hermitean matrices, with the Lebesgue measure:
\beq
dM:= \prod_{i} dM_{ii} \prod_{i<j} d\Re M_{ij} \,d\Im M_{ij}
\eeq

$\bullet$ $GL_n(\C):=$ group of $n\times n$ complex matrices, with the Lebesgue measure:
\beq
dZ:= \prod_{i,j} d\Re Z_{ij} \,d\Im Z_{ij}
\eeq

$\bullet$ $T_n:=$ group of $n\times n$ strictly upper triangular complex matrices, with the Lebesgue measure:
\beq
dT:= \prod_{i<j} d\Re T_{ij} \,d\Im T_{ij}
\eeq

$\bullet$ $D_n(\R):=$ group of $n\times n$ real diagonal matrices, with the Lebesgue measure:
\beq
dX:= \prod_{i} dX_{ii}
\eeq

$\bullet$ $D_n(\C):=$ group of $n\times n$ complex diagonal matrices, with the Lebesgue measure:
\beq
dX:= \prod_{i}  d\Re X_{ii} \,d\Im X_{ii}
\eeq

$\bullet$ $\Sym(n):=$ group of permutations of $n$ elements.

\subsection{Vandermonde determinant}

For any diagonal matrix $X=\diag(X_1,\dots,X_n)\in D_n(\C)$, one writes:
\beq
\Delta(X):=\prod_{i<j} (X_i-X_j)
\eeq
and, for any permutation $\sigma\in\Sym(n)$, we define the diagonal matrix:
\beq
X_\sigma := \diag(X_{\sigma(1)},\dots,X_{\sigma(n)})
\eeq
Notice that:
\beq
\Delta(X_\sigma):= (-1)^\sigma\,\,\Delta(X)
\eeq

\subsection{Invariant functions}

\bd
$F(A,B)$ defined  on $GL_n(\C)\times GL_n(\C)\to \C$  is an analytical invariant function  if:

$\bullet$ $F$ is analytical in each variable,

$\bullet$ $\forall U\in GL_n^*(\C),  F(UAU^{-1},UBU^{-1})=F(A,B)$.
\ed

Examples:
\beq
F(A,B) = \prod_{t=1}^p \, \Tr \left(\prod_{r_t=1}^{R_t} \,( x_{t,r_t}-A)(y_{t,r_t}-B)\right)
\eeq
\beq
F(A,B) = \ee{-\Tr V_1(A)}\,\ee{-\Tr V_2(B)}
\eeq

\bd
Monomial invariant functions are functions of the form:
\beq
F(A,B) = \prod_{t=1}^p \, \Tr \left(\prod_{r_t=1}^{R_t} (A^{k_{t,r_t}}\,B^{l_{t,r_t}})\right)
\eeq
where the $k_{t,r_t}$'s and $l_{t,r_t}$'s are integers such that $k_{t,r_t}+l_{t,r_t}>0$.
The total degree is
\beq
\deg F := \sum_{t=1}^p \sum_{r_t=1}^{R_t}\, k_{t,r_t}+l_{t,r_t}
\eeq
\ed

\bd
Polynomial invariant functions are finite complex linear combinations of monomial invariant functions.
\ed

Examples of polynomial invariant functions:
\beq
F(A,B) = \Tr A^{k_1}\,B^{l_1}\,A^{k_2}\,B^{l_2}\,
\virg
F(A,B) = \left(1+\Tr A^{k_1}\,B^{l_1}\right)\,\left(1+\Tr A^{k_2}\,B^{l_2}\right)
\eeq
\beq
F(A,B) = \prod_{t=1}^p \det(x_t-A)^{k_t} \prod_{u=1}^q \det(y_u-B)^{l_u}
\eeq
\beq
F(A,B) = \det(A\otimes 1 - 1\otimes B)
\eeq

\subsection{Decompositions}

\subsubsection{Diagonalization}

It is a standard result in algebra (see \cite{Mehta, fulton, Knapp} for instance), that any hermitean matrix $M\in  H_n$ can be written:
\beq
M = U X U^\dagger
\eeq
where $U\in U(n)$ and $X\in D_n(\R)$.

The measure is then:
\beq\label{diagoMH}
dM = \td{J}_n\,\Delta^2(X) \, dU\, dX
\eeq
where the Jacobian is
\beq
\td{J}_n={\pi^{n(n-1)\over 2}\over \prod_{k=0}^{n-1} k!}
\eeq
This decomposition is not unique. It is unique up to a permutation of eigenvalues,
and up to multiplication of $U$ by a diagonal matrix whose elements are on the unit circle.
In other words, $M=UXU^\dagger$ provides a mapping between $H_n$ and
$U(n)\times D_n(\R)/(U(1)^n\times \Sym(n))$.

\subsubsection{Jordanization}

A less standard result (see \cite{Mehta, Ginibre, Knapp, fulton} for instance),
is that any complex matrix $Z\in  GL_n(\C)$ can be written:
\beq\label{JordanMC}
Z = U (X+T) U^\dagger
\eeq
where $U\in U(n)$, $T\in T_n$ and $X\in D_n(\C)$.

The measure is then:
\beq
dZ = J_n\, |\Delta(X)|^2 \, dU\, dT\, dX
\eeq
where the Jacobian is
\beq
J_n={\left({\pi\over 2}\right)^{n(n-1)\over 2}\over \prod_{k=0}^{n-1} k!}
\eeq
This decomposition is not unique. It is unique up to a permutation of eigenvalues,
and up to multiplication of $U$ by a diagonal matrix whose elements are on the unit circle.
In other words, $Z=U(X+T)U^\dagger$ provides a mapping between $Gl_n(\C)$ and
$U(n)\times T_n\times D_n(\C)/(U(1)^n\times \Sym(n))$.

\section{Gaussian matrix integrals}

In all what follows, we consider 3 complex numbers $\alpha_1$, $\alpha_2$ and $\gamma$, and we define
\beq
\delta := \alpha_1 \alpha_2-\gamma^2
\eeq
and assume that $\delta\neq 0$.

\subsection{Gaussian Hermitean model}

Consider the measure on $H_n\times H_n$:
\beq
\ee{-\Tr\left({\alpha_1\over 2} M_1^2 +{\alpha_2\over 2} M_2^2 + \gamma M_1 M_2\right)}\, dM_1 \, dM_2
\eeq

\bd
The partition function is:
\beq
Z_H(n,\gamma,\alpha_1,\alpha_2) : =\int_{H_n\times H_n} dM_1\, dM_2\, \ee{-\Tr ({\alpha_1\over 2} M_1^2+{\alpha_2\over 2} M_2^2+\gamma M_1 M_2)}
\eeq
\ed
Notice that the integral $Z_H$ is absolutely convergent only if
\beq\label{condconvH}
\forall \phi\in\R\qquad \Re(\alpha_1 \ee{\phi}+\alpha_2 \ee{-\phi}\pm 2\gamma)>0
\eeq
which implies that $\Re\alpha_1>0$, $\Re\alpha_2>0$, $(\Re\gamma)^2<\Re\alpha_1\Re\alpha_2$.

An easy gaussian integral computation gives:
\beq
Z_H = 2^{n} \left({\pi\over \sqrt\delta}\right)^{n^2} \ .
\eeq

\bd
The expectation value of an invariant function $F(A,B)$ is:
\beq
\left<F\right>_H : =
{\int_{H_n\times H_n} dM_1\, dM_2\,F(M_1,M_2)\, \ee{-\Tr ({\alpha_1\over 2} M_1^2+{\alpha_2\over 2} M_2^2+\gamma M_1 M_2)} \over
\int_{H_n\times H_n} dM_1\, dM_2\, \ee{-\Tr ({\alpha_1\over 2} M_1^2+{\alpha_2\over 2} M_2^2+\gamma M_1 M_2)} }
\eeq
\ed

\br
It is clear, from Wick's theorem, that if $F$ is a monomial invariant function, then $<F>_H$ is a polynomial in ${\alpha_1\over \delta}$, ${\alpha_2\over \delta}$ and ${\gamma\over \delta}$,
and can be analiticaly continued to every complex $\alpha_1,\alpha_2, \gamma$, provided that $\delta\neq 0$.
\er

\subsubsection{Gaussian Hermitean loop equations}

Consider a monomial matrix valued function, of the form:
\beq\label{floopeqH}
f(A,B) = f_0(A,B)\, \prod_{t=1}^p \Tr f_t(A,B)
\virg
\forall t=0,\dots,p,\quad f_t(A,B)=  \prod_{r_t=1}^{R_t} A^{k_{t,r_t}} {B}^{l_{t,r_t}}
\eeq
define:
\beq
G_0(A,B) := \prod_{u\neq 0} \Tr f_u(A,B)
\virg {\rm and\,\, if\,\,} t\geq 1\, , \,\,
G_t(A,B) := \prod_{u\neq 0,t} \Tr f_u(A,B)
\eeq

\bt\label{loopeqH}
One has the ''loop equations'':
\bea\label{loopeqHun}
&&\alpha_1\left<G_0(M_1,M_2)\Tr M_1 f_0(M_1,M_2)\right>_H  + \gamma \left<G_0(M_1,M_2)\Tr M_2 f_0(M_1,M_2)) \right>_H \cr
&=& \sum_{r=1}^{R_0} \sum_{j=0}^{k_{0,r}-1}
\left<G_0(M_1,M_2)\,\Tr\left(\left(\prod_{u=1}^{r-1} M_1^{k_{0,u}} {M_2}^{l_{0,u}}\right) M_1^{j}\right)\, \right. \cr
&& \qquad\qquad\qquad\qquad\left. \Tr\left(M_1^{k_{0,r}-j-1} {M_2}^{l_{0,r}} \left(\prod_{u=r+1}^{R_0}  M_1^{k_{0,u}} {M_2}^{l_{0,u}}\right)\right) \right>_H \cr
&& + \sum_{t=1}^p \sum_{r=1}^{R_t} \sum_{j=0}^{k_{t,r}-1}
 \left<G_t(M_1,M_2)\,\Tr\left(\left(\prod_{u=1}^{r-1} M_1^{k_{t,u}} {M_2}^{l_{t,u}}\right) M_1^{j}f_0(M_1,M_2) M_1^{k_{t,r}-j-1} {M_2}^{l_{t,r}} \right.\right. \cr
&&  \qquad \qquad \qquad\qquad\left.\left. \left(\prod_{u=r+1}^{R_t}  M_1^{k_{t,u}} {M_2}^{l_{t,u}}\right)\right) \right>_H \cr
\eea
and
\bea\label{loopeqHdeux}
&&\alpha_2\left<G_0(M_1,M_2)\Tr M_2 f_0(M_1,M_2)\right>_H  + \gamma \left<G_0(M_1,M_2)\Tr M_1 f_0(M_1,M_2)) \right>_H \cr
&=& \sum_{r=1}^{R_0} \sum_{j=0}^{l_{0,r}-1}
\left<G_0(M_1,M_2)\,\Tr\left(\left(\prod_{u=1}^{r-1} M_1^{k_{0,u}} {M_2}^{l_{0,u}}\right) M_1^{k_{0,r}} {M_2}^{j}\right)\,\right. \cr
&& \qquad\qquad\qquad\qquad\left. \Tr\left({M_2}^{l_{0,r}-j-1} \left(\prod_{u=r+1}^{R_0}  M_1^{k_{0,u}} {M_2}^{l_{0,u}}\right)\right) \right>_H \cr
&& + \sum_{t=1}^p \sum_{r=1}^{R_t} \sum_{j=0}^{l_{t,r}-1}
 \left<G_t(M_1,M_2)\,\Tr\left(\left(\prod_{u=1}^{r-1} M_1^{k_{t,u}} {M_2}^{l_{t,u}}\right) {M_1}^{k_{t,r}} {M_2}^{j}f_0(M_1,M_2) {M_2}^{l_{t,r}-j-1} \right.\right. \cr
&&  \qquad \qquad \qquad\qquad\left.\left.  \left(\prod_{u=r+1}^{R_t}  M_1^{k_{t,u}} {M_2}^{l_{t,u}}\right)\right) \right>_H \cr
\eea
\et
Notice that the RHS is a linear combination of invariant polynomial functions of degree strictly lower than the LHS.

Loop equations are a standard method for finding recursion relations among expectation values \cite{ZJDFG}, they were first studied by \cite{Staudacher} for the 2-matrix model,
and solved more explicitely by \cite{Eyn2mat, eynardchain, Eynmultimat}.

\proof{
Write that the integral of a total derivative is zero:
\beq
0= \sum_i \int dM_1\, dM_2\,{\d\over \d {M_1}_{ii}}\,\left(f_{i,i}(M_1,M_2) \, \ee{-\Tr ({\alpha_1\over 2} M_1^2+{\alpha_2\over 2} M_2^2+\gamma M_1 M_2)}\right)
\eeq
i.e.
\bea\label{loopHderivdiag}
&& \sum_i \int dM_1\, dM_2\,\left({\d\over \d {M_1}_{ii}}\,f_{i,i}(M_1,M_2)\right) \, \ee{-\Tr ({\alpha_1\over 2} M_1^2+{\alpha_2\over 2} M_2^2+\gamma M_1 M_2)} \cr
&=& \sum_i \int dM_1\, dM_2\,f_{i,i}(M_1,M_2) \, \left(\alpha_1 {M_1}_{ii}+\gamma {M_2}_{ii}\right)\,\ee{-\Tr ({\alpha_1\over 2} M_1^2+{\alpha_2\over 2} M_2^2+\gamma M_1 M_2)} \cr
\eea
Similarly:
\bea\label{loopHderivRe}
&& \sum_{i<j} \int dM_1\, dM_2\,\left({\d\over \d \Re{M_1}_{ij}}\,f_{i,j}(M_1,M_2)\right) \, \ee{-\Tr ({\alpha_1\over 2} M_1^2+{\alpha_2\over 2} M_2^2+\gamma M_1 M_2)} \cr
&=& \sum_{i<j} \int dM_1\, dM_2\,f_{i,j}(M_1,M_2) \, \left(\alpha_1 ({M_1}_{ji}+{M_1}_{ij}) +\gamma({M_2}_{ji}+{M_2}_{ij})\right)\cr
&&\qquad\qquad\qquad\qquad\qquad\qquad\,\ee{-\Tr ({\alpha_1\over 2} M_1^2+{\alpha_2\over 2} M_2^2+\gamma M_1 M_2)} \cr
\eea
and
\bea\label{loopHderivim}
&& \sum_{i<j} \int dM_1\, dM_2\,\left({\d\over \d \Im{M_1}_{ij}}\,f_{i,j}(M_1,M_2)\right) \, \ee{-\Tr ({\alpha_1\over 2} M_1^2+{\alpha_2\over 2} M_2^2+\gamma M_1 M_2)} \cr
&=& i \sum_{i<j} \int dM_1\, dM_2\,f_{i,j}(M_1,M_2) \, \left(\alpha_1 ({M_1}_{ji}-{M_1}_{ij}) +\gamma({M_2}_{ji}-{M_2}_{ij})\right)\,\cr
&& \qquad\qquad\qquad\qquad\qquad\ee{-\Tr ({\alpha_1\over 2} M_1^2+{\alpha_2\over 2} M_2^2+\gamma M_1 M_2)} \cr
\eea
Taking \ref{loopHderivdiag} $+$ \ref{loopHderivRe} $-i$ \ref{loopHderivim} , we get:
\bea\label{loopHderiv}
 \sum_{i} \int dM_1\, dM_2\,\left({\d\over \d {M_1}_{ii}}\,f_{i,i}(M_1,M_2)\right) \, \ee{-\Tr ({\alpha_1\over 2} M_1^2+{\alpha_2\over 2} M_2^2+\gamma M_1 M_2)} \cr
 +{1\over 2}\sum_{i<j} \int dM_1\, dM_2\,\left(\left({\d\over \d \Re{M_1}_{ij}}-i{\d\over \d \Im{M_1}_{ij}}\right)\,f_{i,j}(M_1,M_2)\right) \, \ee{-\Tr ({\alpha_1\over 2} M_1^2+{\alpha_2\over 2} M_2^2+\gamma M_1 M_2)} \cr
 +{1\over 2}\sum_{i<j} \int dM_1\, dM_2\,\left(\left({\d\over \d \Re{M_1}_{ij}}+i{\d\over \d \Im{M_1}_{ij}}\right)\,f_{j,i}(M_1,M_2)\right) \, \ee{-\Tr ({\alpha_1\over 2} M_1^2+{\alpha_2\over 2} M_2^2+\gamma M_1 M_2)} \cr
= \int dM_1\, dM_2\,\left(  \Tr f(M_1,M_2) (\alpha_1 {M_1}+\gamma {M_2})\right)\,\ee{-\Tr ({\alpha_1\over 2} M_1^2+{\alpha_2\over 2} M_2^2+\gamma M_1 M_2)} \cr
\eea
i.e. one can proceed as if all the ${M_1}_{ij}$ were $n^2$ real indepedent variables, i.e., by abuse of notation we write:
\bea
&& \sum_{i,j} \int dM_1\, dM_2\,\left({\d\over \d {M_1}_{ij}}\,f_{i,j}(M_1,M_2)\right) \, \ee{-\Tr ({\alpha_1\over 2} M_1^2+{\alpha_2\over 2} M_2^2+\gamma M_1 M_2)} \cr
&=& \int dM_1\, dM_2\,\left(  \Tr  f(M_1,M_2) (\alpha_1 {M_1}+\gamma {M_2})\right)\,\ee{-\Tr ({\alpha_1\over 2} M_1^2+{\alpha_2\over 2} M_2^2+\gamma M_1 M_2)} \cr
\eea

Now, one can use the following rules:

$\bullet$ split rule:
if $f(M_1,M_2) = A M_1^k B$ (where $A$ and $B$ are matrices), one has:
\beq
\sum_{i,j} {\d f(M_1,M_2)_{ij}\over \d {M_1}{ij}} = \sum_{l=0}^{k-1} \Tr\left( A M_1^{k-1-l}\right)\,\Tr\left( M_1^{l} B\right)
\eeq

$\bullet$ merge rule:
if $f(M_1,M_2) = A \Tr (M_1^k B)$ (where $A$ and $B$ are matrices), one has:
\beq
\sum_{i,j} {\d f(M_1,M_2)_{ij}\over \d {M_1}{ij}} = \sum_{l=0}^{k-1} \Tr\left( A M_1^{k-1-l} B M_1^{l} \right)
\eeq

Then, if $A$ and $B$ depend on $M_1$, one has to use the chain rule.

When one considers $f$ given by \ref{floopeqH}, one gets eq.\ref{loopeqHun}.

Eq.\ref{loopeqHdeux} is obtained by doing the same for $M_2$.~}

We find again that $\left<F\right>_H$ is a polynomial in ${\alpha_1\over \delta}$, ${\alpha_2\over \delta}$ and ${\gamma\over \delta}$.

\subsection{Gaussian Complex model}

Consider the measure on $GL_n(\C)$:
\beq
\ee{-\Tr\left({\alpha_1\over 2} Z^2 +{\alpha_2\over 2} {Z^\dagger}^2 + \gamma Z Z^\dagger\right)}\, dZ
\eeq

\bd
The partition function is:
\beq
Z_C(n,\gamma,\alpha_1,\alpha_2) : =\int_{Gl_n(\C)} dZ\, \ee{-\Tr ({\alpha_1\over 2} Z^2+{\alpha_2\over 2} {Z^\dagger}^2+\gamma Z Z^\dagger)}
\eeq
\ed
Notice that the integral $Z_C$ is absolutely convergent only if
\beq\label{condconvC}
\forall \theta\in\R\qquad \Re(\alpha_1 \ee{i\theta}+\alpha_2 \ee{-i\theta}+ 2\gamma)>0
\eeq
One can see that with $\theta=\pi$, this condition can never be compatible with \ref{condconvH} (with $\phi=0$).
Therefore, if $Z_H$ is an absolutely convergent integral then $Z_C$ is not, and vice--versa.

An easy gaussian integration gives (where $\delta = \alpha_1\alpha_2-\gamma^2$):
\beq
Z_C = \left({\pi\over \sqrt{-\delta}}\right)^{n^2}
\eeq
which can be analiticaly continued to every $\alpha_1,\alpha_2, \gamma$, provided that $\delta\neq 0$.

\bd
The expectation value of an invariant function $F(A,B)$ is:
\beq
\left<F\right>_C : =
{\int_{Gl_n(\C)} dZ\, F(Z,Z^\dagger)\,\ee{-\Tr ({\alpha_1\over 2} Z^2+{\alpha_2\over 2} {Z^\dagger}^2+\gamma Z Z^\dagger)} \over
\int_{Gl_n(\C)} dZ \,\ee{-\Tr ({\alpha_1\over 2} Z^2+{\alpha_2\over 2} {Z^\dagger}^2+\gamma Z Z^\dagger)} }
\eeq
\ed

\br
It is clear, from Wick's theorem, that if $F$ is a monomial invariant function, then $<F>_C$ is a polynomial in ${\alpha_1\over \delta}$, ${\alpha_2\over \delta}$ and ${\gamma\over \delta}$,
and can be analiticaly continued to every complex $\alpha_1,\alpha_2, \gamma$, provided that $\delta\neq 0$.
\er

\subsubsection{Gaussian complex loop equations}

Consider a monomial matrix valued function, of the form:
\beq\label{floopeqC}
f(Z,Z^\dagger) = f_0(Z,Z^\dagger)\, \prod_{t=1}^p \Tr f_t(Z,Z^\dagger)
\virg
f_t(Z,Z^\dagger)=  \prod_{r_t=1}^{R_t} Z^{k_{t,r_t}} {Z^\dagger}^{l_{t,r_t}}
\eeq
define:
\beq
G_0(Z,Z^\dagger) := \prod_{u\neq 0} \Tr f_u(Z,Z^\dagger)
\virg {\rm and\,\, if\,\,} t\geq 1\, , \,\,
G_t(Z,Z^\dagger) := \prod_{u\neq 0,t} \Tr f_u(Z,Z^\dagger)
\eeq

\bt\label{loopeqC}
One has the same loop equations than theorem \ref{loopeqH}, with replacing the subscript $H$ by $C$.
\bea\label{loopeqCun}
&&\alpha_1\left<G_0(Z,Z^\dagger)\Tr Z f_0(Z,Z^\dagger)\right>_C  + \gamma \left<G_0(Z,Z^\dagger)\Tr Z^\dagger f_0(Z,Z^\dagger)) \right>_C \cr
&=& \sum_{r=1}^{R_0} \sum_{j=0}^{k_{0,r}-1}
\left<G_0(Z,Z^\dagger)\,\Tr\left(\left(\prod_{u=1}^{r-1} Z^{k_{0,u}} {Z^\dagger}^{l_{0,u}}\right) Z^{j}\right)\, \right. \cr
&& \qquad\qquad\qquad\qquad\left. \Tr\left(Z^{k_{0,r}-j-1} {Z^\dagger}^{l_{0,r}} \left(\prod_{u=r+1}^{R_0}  Z^{k_{0,u}} {Z^\dagger}^{l_{0,u}}\right)\right) \right>_C \cr
&& + \sum_{t=1}^p \sum_{r=1}^{R_t} \sum_{j=0}^{k_{t,r}-1}
 \left<G_t(Z,Z^\dagger)\,\Tr\left(\left(\prod_{u=1}^{r-1} Z^{k_{t,u}} {Z^\dagger}^{l_{t,u}}\right) Z^{j}f_0(Z,Z^\dagger) Z^{k_{t,r}-j-1} {Z^\dagger}^{l_{t,r}} \right.\right. \cr
&&  \qquad \qquad \qquad\qquad\left.\left. \left(\prod_{u=r+1}^{R_t}  Z^{k_{t,u}} {Z^\dagger}^{l_{t,u}}\right)\right) \right>_C \cr
\eea
and
\bea\label{loopeqCdeux}
&&\alpha_2\left<G_0(Z,Z^\dagger)\Tr Z^\dagger f_0(Z,Z^\dagger)\right>_C  + \gamma \left<G_0(Z,Z^\dagger)\Tr Z f_0(Z,Z^\dagger)) \right>_C \cr
&=& \sum_{r=1}^{R_0} \sum_{j=0}^{l_{0,r}-1}
\left<G_0(Z,Z^\dagger)\,\Tr\left(\left(\prod_{u=1}^{r-1} Z^{k_{0,u}} {Z^\dagger}^{l_{0,u}}\right) Z^{k_{0,r}} {Z^\dagger}^{j}\right)\,\right. \cr
&& \qquad\qquad\qquad\qquad\left. \Tr\left({Z^\dagger}^{l_{0,r}-j-1} \left(\prod_{u=r+1}^{R_0}  Z^{k_{0,u}} {Z^\dagger}^{l_{0,u}}\right)\right) \right>_C \cr
&& + \sum_{t=1}^p \sum_{r=1}^{R_t} \sum_{j=0}^{l_{t,r}-1}
 \left<G_t(Z,Z^\dagger)\,\Tr\left(\left(\prod_{u=1}^{r-1} Z^{k_{t,u}} {Z^\dagger}^{l_{t,u}}\right) {Z}^{k_{t,r}} {Z^\dagger}^{j}f_0(Z,Z^\dagger) {Z^\dagger}^{l_{t,r}-j-1} \right.\right. \cr
&&  \qquad \qquad \qquad\qquad\left.\left.  \left(\prod_{u=r+1}^{R_t}  Z^{k_{t,u}} {Z^\dagger}^{l_{t,u}}\right)\right) \right>_C \cr
\eea
\et
Notice that the RHS is a linear combination of invariant polynomial functions of degree strictly lower than the LHS.

\proof{
The proof is very similar to that of theorem \ref{loopeqH}.
Write that the integral of a total derivative is zero:
\bea
0&=&\int dZ {\d\over \d \Re Z_{ij}}\left( f_{r,s}(Z,Z^\dagger)\, \ee{-\Tr({\alpha_1\over 2} Z^2+{\alpha_2\over 2} {Z^\dagger}^2+\gamma Z Z^\dagger)} \right)  \cr
\eea
and
\bea
0&=&-i\int dZ {\d\over \d \Im Z_{ij}}\left( f_{r,s}(Z,Z^\dagger)\, \ee{-\Tr({\alpha_1\over 2} Z^2+{\alpha_2\over 2} {Z^\dagger}^2+\gamma Z Z^\dagger)} \right)  \cr
\eea
Taking the sum of both lines, one can proceed as if all the ${Z}_{ij}$ and ${Z^\dagger}_{ij}$ were real indepedent variables, and from there, follow the proof of theorem \ref{loopeqH}.
}

\br\label{remidZHZC}
We see that the loop equations of both models are identical.
It is clear from the above derivation that this is general, even for non gaussian measures.
When the measure is gaussian, the loop equations determine completely every expectation value, while for non-gaussian measures,
 the loop equations give recursion relations for expectation values, but don't give the initial conditions.

Let us consider in particular the ''semi-classical case'' \cite{Berto, BEHsemclas}, i.e. with a measure of the type
\beq
\d\mu(M_1,M_2)=
\ee{-\Tr[V_1(M_1)+V_2(M_2)+M_1 M_2]}
\eeq
where $V'_1$ and $V'_2$ are rational functions.
In that case, the initial conditions which allow to determine all polynomial expectation values recursively, are in one--to--one correspondance with
homology classes of integration paths for pairs of eigenvalues \cite{BEHsemclas}, therefore, there exists a choice of integration path $\Gamma$
such that one can write:
\beq
\int_{(H_n\times H_n)(\Gamma)} dM_1\, dM_2\, \ee{-\Tr[V_1(M_1)+V_2(M_2)+M_1 M_2]}
\equiv
\int_{GL_n(\C)} dZ\, \ee{-\Tr[V_1(Z)+V_2(Z^\dagger)+Z Z^\dagger]}
\eeq
and one can consider that this equality defines the RHS.
Somehow, the complex matrix model is nothing but the analytical continuation of the 2-matrix model defined on some classes of contours.
\er

\subsection{Relation between the two models}\label{secrelZHZC}

\bt\label{theoremHC}
For any polynomial invariant function $F(A,B)$, one has:
\beq
\left<F\right>_H = \left<F\right>_C
\eeq
\et

Notice that $\left<F\right>_H$ and $\left<F\right>_C$ have been defined for different range of values of $\alpha_1$, $\alpha_2$, and $\gamma$,
but, as we have explained above, both are polynomials of $\alpha_1\over\delta$, $\alpha_2\over\delta$ and $\gamma\over\delta$
(and can be analyticaly continued to any $\alpha_1$, $\alpha_2$, and $\gamma$).
Theorem \ref{theoremHC} is  thus an equality between polynomials.

\proof{
It is sufficient to prove it for monomial invariant functions.
The proof is clearly obtained from the loop equations, by recursion on $\deg F$.
It is obviously true for $\deg F=0$, i.e. $F=1$.
And the loop equations of both models are identical.
}

\bd
For any two given complex diagonal matrices $X$ and $Y$, and any polynomial invariant function $F$, define:
\beq
\td{W}_F(X,Y):=  \Delta^2(X)\,\Delta^2(Y)\,\,\int_{U(n)} dU\,\, F(X,UY U^\dagger)\,\ee{-\gamma \Tr XUY U^\dagger}
\eeq
\beq
\om_F(X,Y):=  \Delta(X)\,\Delta(Y)\,{\int_{T(n)} dT\,\,F(X+T,Y+T^\dagger)\, \ee{-\gamma \Tr T T^\dagger} \over \int_{T(n)} dT\, \ee{-\gamma \Tr T T^\dagger}}
\eeq
which is a polynomial in all its variables $X_i,Y_j$, and a polynomial in $1/\gamma$,
and:
\beq
W_F(X,Y):=  {1\over n!^2}\sum_\sigma\sum_\tau \Delta(X_\sigma)\Delta(Y_\tau)\,\,\ee{-\gamma \Tr X_\sigma Y_\tau}\,\int_{T(n)} dT\,\,F(X_\sigma+T,Y_\tau+T^\dagger)\, \ee{-\gamma \Tr T T^\dagger}
\eeq
\ed

\bt\label{intXYintZ}
For any polynomial invariant function $F(A,B)$, one has:
\bea
&& {\td{J}_n^2\over n!^2\,Z_H(n,\gamma,\alpha_1,\alpha_2)}\,\int_{D_n(\R)\times D_n(R)} dX dY\,\, \ee{-{\alpha_1\over 2}\Tr X^2} \ee{-{\alpha_2\over 2}\Tr Y^2}  \td{W}_F(X,Y)  \cr
&=&{J_n\over n!\,Z_C(n,\gamma,\alpha_1,\alpha_2)}\,\int_{D_n(\C)} dX\,\,  \ee{-{\alpha_1\over 2}\Tr X^2} \ee{-{\alpha_2\over 2}\Tr \ovl{X}^2} {W}_F(X,\ovl{X})
\eea
\et

\proof{
Start from theorem \ref{theoremHC}, diagonalize $M_1$ and $M_2$ on the hermitean side, and jordanize $Z$ on the complex side.
\bea
\left<F\right>_H
&=& {\td{J}_n^2\over n!^2\,Z_H(n,\gamma,\alpha_1,\alpha_2)}\,\int_{D_n(\R)\times D_n(R)} dX dY\,\, \ee{-{\alpha_1\over 2}\Tr X^2} \ee{-{\alpha_2\over 2}\Tr Y^2}  \td{W}_F(X,Y)  \cr
=\left<F\right>_C&=&{J_n\over n!\,Z_C(n,\gamma,\alpha_1,\alpha_2)}\,\int_{D_n(\C)} dX\,\,  \ee{-{\alpha_1\over 2}\Tr X^2} \ee{-{\alpha_2\over 2}\Tr \ovl{X}^2} \ee{-\gamma\Tr X \ovl{X}} \,\om_F(X,\ovl{X}) \cr
&=&{J_n\over n!\,Z_C(n,\gamma,\alpha_1,\alpha_2)}\,\int_{D_n(\C)} dX\,\,  \ee{-{\alpha_1\over 2}\Tr X^2} \ee{-{\alpha_2\over 2}\Tr \ovl{X}^2} \ee{-\gamma\Tr X_\sigma \ovl{X}_\tau} \,\om_F(X_\sigma,\ovl{X}_\tau) \cr
&=&{J_n\over n!\,Z_C(n,\gamma,\alpha_1,\alpha_2)}\,\int_{D_n(\C)} dX\,\,  \ee{-{\alpha_1\over 2}\Tr X^2} \ee{-{\alpha_2\over 2}\Tr \ovl{X}^2}  \,W_F(X_\sigma,\ovl{X}_\tau) \cr
\eea
The equality in the first line is obtained by diagonalizing $M_1$ and $M_2$ (with Jacobian given in eq.\ref{diagoMH}),
the equality in the second line is obtained by Jordanizing $Z$ (with Jacobian given in eq.\ref{JordanMC}),
the equality between the second and third line holds for any pair of permutations $\sigma$ and $\tau$ (it can be proven with the Lemma \ref{lemmaappendix} given in appendix),
and the equality of the last line comes from the definition of $W_F$.
}

\section{Unitary group integrals}

Here is one of the most important theorems of this paper:

\subsection{Unitary integrals and triangular integrals}

\bt\label{ThmintUintT}
For any invariant function $F(A,B)$ one has:
\bea\label{maintheorem}
&&   \int_{U(n)} dU\, F(X,U Y U^\dagger)\, \ee{- \gamma \Tr X U Y U^\dagger}  \cr
&=& {c_n\over n!}\,{\sum_\sigma \sum_\tau (-1)^\sigma (-1)^\tau \ee{-\gamma \Tr X_\sigma Y_\tau} \int_{T(n)} F(X_\sigma+T,Y_\tau+T^\dagger)\, \ee{- \gamma \Tr T T^\dagger}\, dT
\over \Delta(X)\Delta(Y) } \cr
\eea
where
\beq
c_n = {\prod_{k=0}^{n-1} k!\over (-2\pi)^{n(n-1)\over 2}}
\eeq
i.e.
\beq
\,\td{W}_F(X,Y)=n!\,c_n\,{W}_F(X,Y)
\eeq
\et

\proof{
Using the Lemma \ref{lemmaappendix} given in appendix, and using theorem \ref{intXYintZ}, we have:
\bea
&&\int_{D_n(\R)\times D_n(\R)} dX\,dY\,\,\, \ee{-{\alpha_1\over 2}\Tr X^2} \ee{-{\alpha_2\over 2}\Tr Y^2}  \td{W}_F(X,Y) \cr
&=&{n!^2\,Z_H\over \td{J}_n^2}\, {J_n\over n!\,Z_C}\,\int_{D_n(\C)} dX\,\, \, \ee{-{\alpha_1\over 2}\Tr X^2} \ee{-{\alpha_2\over 2}\Tr \ovl{X}^2}  {W}_F(X,\ovl{X}) \cr
&=&{n!^2\,Z_H\over \td{J}_n^2}\,{J_n\over n!\,Z_C}\,{1\over n!^2}\sum_{\sigma,\tau}\,\int_{D_n(\C)} dX\,\, \, \ee{-{\alpha_1\over 2}\Tr X^2} \ee{-{\alpha_2\over 2}\Tr \ovl{X}^2}\ee{-\gamma\Tr X_\sigma\ovl{X}_\tau}  \om_F(X_\sigma,\ovl{X}_\tau) \cr
&=&{n!^2\,Z_H\over \td{J}_n^2\,\left({2\pi\over\sqrt{\delta}}\right)^n}\,{J_n\, \left({\pi\over\sqrt{-\delta}}\right)^n\over n!\,Z_C}\,{1\over n!^2}\sum_{\sigma,\tau}\,\int_{D_n(\R)\times D_n(\R)} \!\!\!\!\!\!\!\!\!\!\!\!\!\!\!\!\!\!
\!dX\,dY\,\, \, \ee{-{\alpha_1\over 2}\Tr X^2} \ee{-{\alpha_2\over 2}\Tr Y^2}\ee{-\gamma\Tr X_\sigma Y_\tau}  \om_F(X_\sigma,Y_\tau) \cr
&=& {n!^2\,Z_H\over \td{J}_n^2\,\left({2\pi\over\sqrt{\delta}}\right)^n}\,{J_n\, \left({\pi\over\sqrt{-\delta}}\right)^n\over n!\,Z_C}\,\int_{D_n(\R)\times D_n(\R)} dX\,dY\,\, \, \ee{-{\alpha_1\over 2}\Tr X^2} \ee{-{\alpha_2\over 2}\Tr Y^2}  W_F(X,Y) \cr
&=& n! c_n\,\int_{D_n(\R)\times D_n(\R)} dX\,dY\,\, \, \ee{-{\alpha_1\over 2}\Tr X^2} \ee{-{\alpha_2\over 2}\Tr Y^2}  W_F(X,Y) \cr
\eea
Notice that if $f(A)$ and $g(B)$ are invariant functions i.e. $f(UAU^{-1})=f(A)$ for all $A$ and $U$ (resp. $g(UBU^{-1})=g(B)$ for all $B$ and $U$), one has:
\beq
W_{f(X)g(Y)F(X,Y)}(X,Y) = f(X)g(Y) W_F(X,Y)
\virg
\td{W}_{f(X)g(Y)F(X,Y)}(X,Y) = f(X)g(Y) \td{W}_F(X,Y)
\eeq

Thus, for any $f$ and $g$:
\beq
0=\int_{D_n(\R)\times D_n(\R)} dX\,dY\,\,  \ee{-{\alpha_1\over 2}\Tr X^2} \ee{-{\alpha_2\over 2}\Tr Y^2} f(X)g(Y) (n!\, c_n\,W_F(X,Y)-\td{W}_F(X,Y))
\eeq
Since  $\td{W}_F(X,Y)$ and $n!c_n{W}_F(X,Y)$ are symmetric functions and entire functions of all their variables, they must be identicaly equal to one another.
}

\subsection{Examples}

Let us illustrate theorem \ref{ThmintUintT} on some simple examples and recover some classical results.

\subsubsection{Harish-Chandra--Itzykson--Zuber's formula}

We can use theorem \ref{ThmintUintT}, to find a new proof of the famous Harish-Chandra--Itzykson--Zuber's formula \cite{IZ, Knapp}.

Indeed, consider $F(A,B)=1$, theorem \ref{ThmintUintT} gives:
\bea
  \int_{U(n)} \, \ee{-\gamma \Tr X U Y U^\dagger}
&=& {c_n\over n!}\,{\sum_{\sigma,\tau} (-1)^\sigma\,(-1)^\tau\,  \prod_i \ee{- \gamma X_{\sigma_i} Y_{\tau_i}} \over \Delta(X)\Delta(Y)  } \,\int_{T_n} dT \, \ee{-\gamma\Tr T T^\dagger} \cr
&=&  c_n\,\left(\pi\over \gamma\right)^{n(n-1)\over 2}\, {\det{E}\over \Delta(X)\Delta(Y)  } \cr
\eea
which is the famous Harish Chandra-Itzykzon-Zuber integral.
Here $E$ is the matrix
\beq
E_{ij} := \ee{-\gamma X_i Y_j}
\eeq

\subsubsection{Morozov's formula}

Consider $\Tr A^k B^l$ for any  integers $k$ and $l$.
It is in fact simpler to introduce a generating function:
\beq
F(A,B)=\Tr {1\over x-A}{1\over y-B} = \sum_{k=0}^\infty \sum_{l=0}^\infty {1\over x^{k+1}}\,{1\over y^{l+1}}\,\Tr A^k B^l
\eeq
which is to be understood as a formal power series in its large $x$ and large $y$ expansion.
$F(A,B)$ is merely a convenient way of considering all polynomial invariant functions of type $\Tr A^k B^l$ at once.

We have:
\beq
{1\over x-(X+T)} = \sum_{p=0}^n \left({1\over x-X}\,T\right)^p\,{1\over x-X}
\eeq
and thus:
\bea
&& {1\over \int_{T(n)}\, dT\,\ee{-\gamma\Tr T T^\dagger}}\,
 \int_{T(n)} \Tr {1\over x-(X+T)}\,{1\over y-(Y+T^\dagger)}\,dT\, \ee{-\gamma\Tr T T^\dagger}  \cr
&=& {1\over \int_{T(n)}\, dT\,\ee{-\gamma\Tr T T^\dagger}}\,\sum_{p=0}^n\sum_{q=0}^n
 \int_{T(n)} \Tr \left({1\over x-X}\,T\right)^p{1\over x-X}\,{1\over y-Y}\left(T^\dagger {1\over y-Y}\right)^q\,dT\, \ee{-\gamma\Tr T T^\dagger}  \cr
&=& \sum_{p=0}^n\sum_{q=0}^n \sum_{i_1<i_2\dots<i_{p+1}}\sum_{j_1<j_2\dots<j_{q+1}}
 \delta_{i_1,j_1}\delta_{i_p,j_q}\,\prod_{k=1}^{p+1}{1\over x-X_{i_k}}\prod_{l=1}^{q+1}{1\over y-Y_{j_l}}\,\cr
&& {\int_{T(n)} T_{i_1,i_2}T_{i_2,i_3}\dots T_{i_p,i_{p+1}} T^\dagger_{j_{q+1},j_q}\dots T^\dagger_{j_2,j_1}\,dT\, \ee{-\gamma\Tr T T^\dagger}\over \int_{T(n)}\, dT\,\ee{-\gamma\Tr T T^\dagger}}
\eea
That last integral is non vanishing only if $p=q$, and according to Wick's theorem, it is the sum of all possible pairings.
Because of the ordering of the $i_k$'s and $j_l$'s, the only non vanishing pairing is obtained for $i_k=j_k$ for all $k$.
Therefore:
\bea
&& {1\over \int_{T(n)}\, dT\,\ee{-\gamma\Tr T T^\dagger}}\,
 \int_{T(n)} \Tr {1\over x-(X+T)}\,{1\over y-(Y+T^\dagger)}\,dT\, \ee{-\gamma\Tr T T^\dagger} \cr
&=& \sum_{p=0}^\infty \sum_{i_1<i_2\dots<i_{p+1}} \,{1\over \gamma^p}\,\prod_{k=1}^{p+1}{1\over (x-X_{i_k}) (y-Y_{i_k})}\,\cr
&=& -\gamma+
\gamma\,\prod_{i=1}^{n}\left(1+{1\over \gamma(x-X_{i}) (y-Y_{i})}\right)\,
\eea

and then theorem \ref{ThmintUintT} gives:
\bea
&&  \left(\gamma\over \pi\right)^{n(n-1)\over 2}\,{\int_{U(n)} dU\,\Tr\left({1\over x-X}U{1\over y-Y} U^\dagger\right)\, \ee{-\gamma \Tr X U Y U^\dagger}} \cr
&=& {c_n\,\gamma\over n!}\,{\sum_{\sigma,\tau} (-1)^\sigma (-1)^\tau\,  \left(-\prod_i \ee{- \gamma X_{\sigma_i} Y_{\tau_i}}+\prod_{i} \left(\ee{- \gamma X_{\sigma_i} Y_{\tau_i}}+{1\over \gamma}\,{1\over x-X_{\sigma_i}}\ee{- \gamma X_{\sigma_i} Y_{\tau_i}}{1\over y-Y_{\tau_i}}\right)\right)
\over \Delta(X)\Delta(Y)  } \cr
&=& \gamma\,c_n\,  {-\det{E}+\det{ \left(E+{1\over \gamma}\,{1\over x-X}E{1\over y-Y}\right)} \over \Delta(X)\Delta(Y)   } \cr
&=&  \gamma \,  \left(-1+\det{ \left(1+{1\over \gamma}\,{1\over x-X}E{1\over y-Y}E^{-1}\right)}\right)\,\, c_n\,{\det{E}\over \Delta(X)\Delta(Y)}
\eea
i.e.
\beq
{\int_{U(n)} dU\,\Tr\left({1\over x-X}U{1\over y-Y} U^\dagger\right)\, \ee{-\gamma \Tr X U Y U^\dagger}
\over \int_{U(n)} dU\, \ee{-\gamma \Tr X U Y U^\dagger}}
=  \gamma \,  \left(-1+\det{ \left(1+{1\over \gamma}\,{1\over x-X}E{1\over y-Y}E^{-1}\right)}\right)
\eeq
which is identical (for $\gamma=-1$) to what was found in \cite{BEmixed, eynmorozov}, i.e. the compact version of Morozov's formula \cite{morozov}.

\section{Computation of triangular integrals}

The goal of this section is to compute the triangular integral on the RHS of theorem \ref{ThmintUintT}.
Here, we consider $\gamma=1$.

\subsection{Parametrization of polynomial invariant functions}

\bd
Let $R$ be a positive integer.
Let $\vec{x}=(x_1,\dots,x_R)$ and $\vec{y}=(y_1,\dots,y_R)$ be $2R$ complex numbers.
Let $\pi$ and $\pi'$ be two permutations of $\Sigma_R$.

The permutation $\pi \pi'^{-1}$ is made of $p$ cycles $C_1,\dots, C_p$ of lenght $R_1,\dots,R_p$ which we note:
\beq
C_k=({i_{k,1}}\mathop\to^{\pi} {j_{k,1}} \mathop\lto^{\pi'^{-1}} {i_{k,2}} \mathop\to^{\pi} {j_{k,2}} \mathop\lto^{\pi'^{-1}} {i_{k,3}} \mathop\to^{\pi}
\dots  \mathop\lto^{\pi'^{-1}} {i_{k,R_k}} \mathop\to^{\pi} {j_{k,R_k}} \mathop\lto^{\pi'^{-1}} {i_{k,1}} )
\eeq
We define, for $(A,B)\in GL_n(\C)^2$ in any dimension $n$:
\beq
F_{\pi,\pi'}(\vec{x},\vec{y},A,B):= \prod_{k=1}^p \left(\delta_{R_k,1}+\Tr \prod_{l=1}^{R_k} {1\over x_{i_{k,l}}-A}{1\over y_{j_{k,l}}-B} \right)
\eeq
As explained above, this definition is to be understood as a formal power series in the large $x_i$ and $y_j$ expansions,
it is merely a way of considering all polynomial invariant functions at once.

\ed

Examples: with $R=2$, we have:
\bea
&& F_{(1)(2),(1)(2)}(x_1,x_2,y_1,y_2,A,B) = \left(1+\Tr {1\over x_1-A}{1\over y_1-B}\right)\,\left(1+\Tr {1\over x_2-A}{1\over y_2-B}\right)  \cr
&& F_{(12),(12)}(x_1,x_2,y_1,y_2,A,B) = \left(1+\Tr {1\over x_1-A}{1\over y_2-B}\right)\,\left(1+\Tr {1\over x_2-A}{1\over y_1-B}\right)  \cr
&& F_{(1)(2),(12)}(x_1,x_2,y_1,y_2,A,B) = \Tr {1\over x_1-A}{1\over y_1-B}{1\over x_2-A}{1\over y_2-B}  \cr
&& F_{(12),(1)(2)}(x_1,x_2,y_1,y_2,A,B) = \Tr {1\over x_1-A}{1\over y_2-B}{1\over x_2-A}{1\over y_1-B}  \cr
\eea

\bd
Let $R$ be a positive integer, $\vec{x}=(x_1,\dots,x_R)$ and $\vec{y}=(y_1,\dots,y_R)$ be $2R$ complex numbers.
Let $\pi$ and $\pi'$ be two permutations of $\Sigma_R$.
Let $n$ be an integer, and $X=\diag(X_1,\dots,X_n)$ and $Y=\diag(Y_1,\dots,Y_n)$ be two complex diagonal matrices of size $n$,
We define:
\beq
W^{(n)}_{\pi,\pi'}(\vec{x},\vec{y},X,Y) := 1  \qquad{\rm if }\,\, n=0\,\, {\rm or}\,\,  R=0
\eeq
\beq
W^{(n)}_{\pi,\pi'}(\vec{x},\vec{y},X,Y) := F_{\pi,\pi'}(\vec{x},\vec{y},X_1,Y_1)\,\,  \qquad{\rm if }\,\, n= 1
\eeq
and otherwise
\beq
W^{(n)}_{\pi,\pi'}(\vec{x},\vec{y},X,Y) :={\int_{T(n)} dT\,\ee{-\Tr T T^\dagger}\, F_{\pi,\pi'}(\vec{x},\vec{y},X+T,Y+T^\dagger)
\over \int_{T(n)} dT\, \ee{-\Tr T T^\dagger}}
\eeq

Here, ${1\over x-(X+T)}$ is defined by:
\beq
\left({1\over x-(X+T)}\right)_{i,j} := {\delta_{ij}\over x-X_i}+\sum_{p=1}^{(j-i)} \sum_{i<i_1<\dots<i_p<j} {1\over x-X_{i}}T_{i,i_1}{1\over x-X_{i_1}}T_{i_1,i_2}\dots {1\over x-X_{i_p}}T_{i_p,j}{1\over x-X_{j}}
\eeq

\ed

\subsection{Computation of triangular integrals of invariant functions}

We are now going to find some recursion relation in $n$ for the $W$'s.

\bt

\beq
W^{(n)}_{\pi,\pi'}(\vec{x},\vec{y},X,Y) = \sum_{\rho} {\cal M}^{(R)}_{\pi,\rho}(\vec{x},\vec{y},X_n,Y_n)\,\, W^{(n-1)}_{\rho,\pi'}(\vec{x},\vec{y},\td{X},\td{Y})
\eeq
where $\td{X}:=\diag(X_1,\dots,X_{n-1})$, $\td{Y}:=\diag(Y_1,\dots,Y_{n-1})$, and:
\beq\label{defMR}
{\cal M}^{(R)}_{\pi,\rho}(\vec{x},\vec{y},X_n,Y_n) = \prod_{i=1}^R \left(\delta_{\pi(i),\rho(i)}+{1\over (x_i-X_n)(y_{\pi(i)}-Y_n)}\right)
\eeq

\et
\proof{
If $T$ is a strictly upper triangular matrix of size $n$, we define $\td{T}$ the triangular matrix of size $n-1$, such that
$\td{T}_{i,j}=T_{i,j}$ for all $i,j<n$, and $\vec{u}$  the vector made of the last column of $T$, $u_k=T_{k,n}$:
\beq
T=\pmatrix{\ddots & \dots &\dots &\vdots & u_1\cr &\ddots & \td{T} &\vdots & \vdots \cr & & \ddots &\vdots & \vdots \cr & & & \ddots & u_{n-1}\cr & & & & 0 }
\eeq
We define $\left({1\over x-(\td{X}+\td{T})}\right)_{i,j}:=0$ if $i=n$ or $j=n$.

Notice that:
\beq\label{xT3terms}
 \left({1\over x-(X+T)}\right)_{i,j}
= \left({1\over x-(\td{X}+\td{T})}\right)_{i,j}  + {\delta_{j,n}\over x-X_n}\sum_{k=1}^{n-1}\left({1\over x-(\td{X}+\td{T})}\right)_{i,k} u_k+ {\delta_{i,n}\delta_{j,n}\over x-X_n}
\eeq
and
\bea\label{xT3termsTr}
&&1+\Tr {1\over x-(X+T)}{1\over y-(Y+T^\dagger)}\cr
&=& 1+\Tr {1\over x-(\td{X}+\td{T})}{1\over y-(\td{Y}+\td{T}^\dagger)} \cr
&& + {1\over (x-X_n)(y-Y_n)}\left(1+\sum_{k=1}^{n-1}\sum_{l=1}^{n-1}\left({1\over y-(\td{Y}+\td{T}^\dagger)}{1\over x-(\td{X}+\td{T})}\right)_{l,k}u_k  \ovl{u}_l \right)  \cr
\eea
Now, we integrate $u$ out, using Wick's theorem, i.e. take the sum over all possible pairings of a $u$ and a $\ovl{u}$.
The pairing $(u_k,\ovl{u}_l)$ gives a factor $\delta_{k,l}$.

Let us represent $W$ as a bivalent graph $G$, whose edges are pairs $(x_i,y_{\pi(i)})$, and whose vertices are pairs $(y_{\pi'(i)},x_i)$.

Relation eq.\ref{xT3terms} means that, for each edge $(x_i,y_{\pi(i)})$ of $G$, we can either:

- let the edge untouched (first term in eq.\ref{xT3terms}), with weight $1$,

- remove the edge (second term in eq.\ref{xT3terms}), with weight ${1\over (x_i-X_n)(y_{\pi(i)}-Y_n)}$,

- remove the vertex $(y_{\pi'(i)},x_i)$ (third term in eq.\ref{xT3terms}), with weight ${1\over (x_i-X_n)(y_{\pi'(i)}-Y_n)}$, which means that either the neighboring edges cannot stay untouched.

Then, we integrate $u$ out, i.e. we take the sum over all possible pairings, i.e. we draw new edges between vertices (those not removed), so that the final graph is bivalent.
For each pairing, we get a new graph $G'$.
The sum over possible pairings, is thus the sum over bivalent graphs $G'$, whose vertices form a subset of the vertices of $G$, i.e.
\beq
W_{G}^{(n)} = \sum_{G'} {\cal M}_{G,G'}\,W_{G'}^{(n-1)}
\eeq
where the coefficient ${\cal M}_{G,G'}$ is computed as follows:

- ${\cal M}_{G,G'}$ receives a factor $1+{1\over (x_i-X_n)(y_{\pi(i)}-Y_n)}$ for each edge $(x_i,y_{\pi(i)})$ of $G$ which is unchanged, i.e. which is an edge of $G'$.
($1$ if it was not removed, and ${1\over (x_i-X_n)(y_{\pi(i)}-Y_n)}$ if it was  removed and drawn again).

- the weight of each edge $(x_i,y_{\pi(i)})$ of $G$, which is not an edge of $G'$, is ${1\over (x_i-X_n)(y_{\pi(i)}-Y_n)}$.

- the weight of removing a vertex is the same as the weight of creating a lenght $1$ cycle at that vertex.
In other words, if $G'$ has less vertices than $G$, consider $G''$ obtained from $G'$ by adding lenght $1$ cycles at each missing vertex, one has
${\cal M}_{G,G'}={\cal M}_{G,G''}$.
The sum over $G'$ can thus be written as a sum over $G''$, where $G''$ has as many vertices as $G$, and all cycles of lenght $1$ come together with a $1$ added.

- relation eq.\ref{xT3termsTr} ensures that the previous rules apply also when $G$ has lenght $1$ cycles.

To sumarize, we have:
\beq
W_{G}^{(n)} = \sum_{G''} {\cal M}_{G,G''}\,W_{G''}^{(n-1)}
\eeq
where
\beq
{\cal M}_{G,G''} =
\prod_{(x_i,y_{\pi(i)})\in G''} \left(1+{1\over (x_i-X_n)(y_{\pi(i)}-Y_n)}\right)
\,\prod_{(x_i,y_{\pi(i)})\notin G''} {1\over (x_i-X_n)(y_{\pi(i)}-Y_n)}
\eeq

when $G$ and $G''$ are written in terms of pairs of permutations, it reduces to eq.\ref{defMR}.
}

\bigskip

\br
Notice that:
\beq
{\cal M}^{(R)}(\vec{x},\vec{y},X_n,Y_n)={\cal M}^{(R)}(\vec{x},\vec{y},X_n,Y_n)^t
\eeq
\beq
{\cal M}^{(R)}_{\pi,\pi'}(\vec{x},\vec{y},X_n,Y_n) = {\cal M}^{(R)}_{\pi^{-1},\pi'^{-1}}(\vec{y},\vec{x},Y_n,X_n)
\eeq
\beq
{\cal M}^{(R)}_{\pi\rho,\pi'\rho}(\vec{x},\vec{y},X_n,Y_n) = {\cal M}^{(R)}_{\pi,\pi'}(\vec{x}_{\rho^{-1}},\vec{y},X_n,Y_n)
\eeq
\er

\bt\label{ThmWPiM}
\beq
W^{(n)}_{\pi,\pi'}(\vec{x},\vec{y},X,Y)
= \left({\cal M}^{(R)}(\vec{x},\vec{y},X_n,Y_n)\, {\cal M}^{(R)}(\vec{x},\vec{y},X_{n-1},Y_{n-1})\, \dots  {\cal M}^{(R)}(\vec{x},\vec{y},X_1,Y_1)       \right)_{\pi,\pi'}
\eeq
\et

\proof{
For $n=1$, we have
\beq
W^{(1)}_{\pi,\pi'}(\vec{x},\vec{y},X_1,Y_1) = F_{\pi,\pi'}(\vec{x},\vec{y},X_1,Y_1)
= {\cal M}^{(R)}_{\pi,\pi'}(\vec{x},\vec{y},X_1,Y_1)
\eeq
The proof follows from recursion on $n$.
}

\bt\label{thmMcommute}
The matrices ${\cal M}^{(R)}(\vec{x},\vec{y},\xi,\eta)$ commute among themselves:
\beq
{\cal M}^{(R)}(\vec{x},\vec{y},\xi,\eta){\cal M}^{(R)}(\vec{x},\vec{y},\xi',\eta')
={\cal M}^{(R)}(\vec{x},\vec{y},\xi',\eta'){\cal M}^{(R)}(\vec{x},\vec{y},\xi,\eta)
\eeq
\et
\proof{
Let $n=2$, $X=\diag(X_1,X_2)$ and $Y=\diag(Y_1,Y_2)$ be two diagonal matrices, and
$\td{X}=\diag(X_2,X_1)$ and $\td{Y}=\diag(Y_2,Y_1)$.
Let $T$ be a $2\times 2$ upper triangular matrix with non vanishing element $T_{12}$.
Let $U$ be the $2\times 2$ matrix:
\beq
U=\pmatrix{ \ovl{T}_{12} & Y_2-Y_1 \cr X_1-X_2 & T_{12}}
\eeq
it satisfies:
\beq
U(X+T)=(\td{X}+\ovl{T})U
\virg
U(Y+T^\dagger)=(\td{Y}+T^t)U
\eeq
If $U$ is invertible (which is true for almost every $T$), one has:
\beq
F_{\pi,\rho}(\vec{x},\vec{y},X+T,Y+T^\dagger) = F_{\pi,\rho}(\vec{x},\vec{y},\td{X}+\ovl{T},\td{Y}+T^t)
\eeq
for every $T$ (except a zero measure subset).
Since the Jacobian $\left|{\d \ovl{T}\over \d T}\right|=1$, one has:
\bea\label{intTpermuted}
 {\int_{T(2)} dT\,\ee{-\Tr T T^\dagger}\, F_{\pi,\rho}(\vec{x},\vec{y},X+T,Y+T^\dagger)
\over \int_{T(2)} dT\, \ee{-\Tr T T^\dagger}}
=
{\int_{T(2)} d\td{T}\,\ee{-\Tr \td{T} \td{T}^\dagger}\, F_{\pi,\rho}(\vec{x},\vec{y},\td{X}+\ovl{T},\td{Y}+T^t)
\over \int_{T(2)} d\td{T}\, \ee{-\Tr \td{T} \td{T}^\dagger}}
\eea
Using theorem \ref{ThmWPiM} for $n=2$, we have:
\beq
{\cal M}^{(R)}(\vec{x},\vec{y},X_1,Y_1) {\cal M}^{(R)}(\vec{x},\vec{y},X_2,Y_2)
=
{\cal M}^{(R)}(\vec{x},\vec{y},X_2,Y_2) {\cal M}^{(R)}(\vec{x},\vec{y},X_1,Y_1)
\eeq
}

\bc
therefore, there exists an orthogonal matrix ${\cal U}(\vec{x},\vec{y})$, independent of $\xi$ and $\eta$, such that:
\beq
\L(\vec{x},\vec{y},\xi,\eta)\,
:={\cal U}(\vec{x},\vec{y})\,{\cal M}^{(R)}(\vec{x},\vec{y},\xi,\eta)\,{\cal U}^t(\vec{x},\vec{y})
\eeq
is a diagonal matrix
\beq
\L(\vec{x},\vec{y},\xi,\eta) = \diag\left(\L_\pi(\vec{x},\vec{y},\xi,\eta)\right)
\eeq
\ec
Notice that $\L$ is a rational function of $\xi$ and $\eta$.

Thus:
\beq
{\int_{T(n)} dT\,\ee{-\Tr T T^\dagger}\, F_{\pi,\pi'}(\vec{x},\vec{y},X+T,Y+T^\dagger)
\over \int_{T(n)} dT\, \ee{-\Tr T T^\dagger}}
= \sum_\rho {\cal U}_{\pi,\rho}(\vec{x},\vec{y})\,{\cal U}_{\pi',\rho}(\vec{x},\vec{y})   \,\prod_{i=1}^n \L_\rho(\vec{x},\vec{y},X_i,Y_i)
\eeq
and:
\beq\label{UnintLambdaperm}
{\int_{U(n)} dU\,\ee{-\Tr XUYU^\dagger}\, F_{\pi,\pi'}(\vec{x},\vec{y},X,UYU^\dagger)
\over \int_{U(n)} dU\,\ee{-\Tr XUYU^\dagger}}
= \sum_\rho {\cal U}_{\pi,\rho}(\vec{x},\vec{y})\,{\cal U}_{\pi',\rho}(\vec{x},\vec{y})   {\det\left(\ee{-X_i Y_j}\L_\rho(\vec{x},\vec{y},X_i, Y_j)\right)\over \det\left(\ee{-X_i Y_j}\right)}
\eeq

\br
if one defines the ''Matricial determinant'' as follows:
\bd\label{defMdet}
Let $M\in GL_n(Gl_m(\C))$, i.e. for each $i=1,\dots,n$, $j=1,\dots,n$, $M_{i,j}$ is a square matrices of size $m$.
We define:
\beq
\Mdet(M) := {1\over n!}\,\sum_{\sigma\in \Sym(n)}\sum_{\tau\in \Sym(n)} (-1)^\sigma(-1)^\tau \prod_{i=1}^n M_{\sigma(i),\tau(i)}
\eeq
which is a $m\times m$ square matrix.
\ed
then we have:
\beq
\int_{U(n)} dU\, F_{\pi,\pi'}(\vec{x},\vec{y},X,UYU^\dagger)\,\ee{-\Tr XUYU^\dagger}
= c_n\,\left(\pi\right)^{n(n-1)\over 2}\,{ \left(\Mdet{\left( \ee{-X_i Y_j}\,{\cal M}^{(R)}(\vec{x},\vec{y},X_i,Y_j)\right)}\right)_{\pi,\pi'}\over \Delta(X)\Delta(Y)}
\eeq
if $R=0$, one immediately recovers the Itzykson--Zuber's formula, and if $R=1$, one immediately recovers Morozov's formula.
\er

\subsection{Examples}

$\bullet$ Example $R=1$:
\beq
{\cal M}^{(1)}_{\1,\1}(x,y,\xi,\eta) = 1+{1\over x-\xi}{1\over y-\eta}
\eeq
and thus:
\beq
{\int_{T(n)} dT\,\ee{-\Tr T T^\dagger}\, \left(1+\Tr {1\over x-(X+T)}{1\over x-(Y+T^\dagger)}\right)
\over \int_{T(n)} dT\, \ee{-\Tr T T^\dagger}}
= \prod_{i=1}^n \left(1+{1\over x-X_i}{1\over y-Y_i}\right)
\eeq

$\bullet$ Example $R=2$:

We have:
\bea
&& F_{(1)(2),(1)(2)}(x_1,x_2,y_1,y_2,A,B) = \left(1+\Tr {1\over x_1-A}{1\over y_1-B}\right)\,\left(1+\Tr {1\over x_2-A}{1\over y_2-B}\right)  \cr
&& F_{(12),(12)}(x_1,x_2,y_1,y_2,A,B) = \left(1+\Tr {1\over x_1-A}{1\over y_2-B}\right)\,\left(1+\Tr {1\over x_2-A}{1\over y_1-B}\right)  \cr
&& F_{(1)(2),(12)}(x_1,x_2,y_1,y_2,A,B) = \Tr {1\over x_1-A}{1\over y_1-B}{1\over x_2-A}{1\over y_2-B}  \cr
&& F_{(12),(1)(2)}(x_1,x_2,y_1,y_2,A,B) = \Tr {1\over x_1-A}{1\over y_2-B}{1\over x_2-A}{1\over y_1-B}  \cr
\eea
and
\beq
\left\{
\begin{array}{l}
{\cal M}^{(2)}_{(1)(2),(1)(2)}(x_1,x_2,y_1,y_2,\xi,\eta) = \left(1+{1\over x_1-\xi}{1\over y_1-\eta}\right)\left(1+{1\over x_2-\xi}{1\over y_2-\eta}\right)  \cr
{\cal M}^{(2)}_{(12),(12)}(x_1,x_2,y_1,y_2,\xi,\eta) = \left(1+{1\over x_1-\xi}{1\over y_2-\eta}\right)\left(1+{1\over x_2-\xi}{1\over y_1-\eta}\right)  \cr
{\cal M}^{(2)}_{(1)(2),(12)}(x_1,x_2,y_1,y_2,\xi,\eta) ={1\over x_1-\xi}{1\over y_1-\eta}{1\over x_2-\xi}{1\over y_2-\eta}  \cr
{\cal M}^{(2)}_{(12),(1)(2)}(x_1,x_2,y_1,y_2,\xi,\eta) ={1\over x_1-\xi}{1\over y_2-\eta}{1\over x_2-\xi}{1\over y_1-\eta}
\end{array}
\right.
\eeq
i.e., the matrix ${\cal M}^{(2)}(x_1,x_2,y_1,y_2,\xi,\eta)$ is:
\beq
\pmatrix{
\left(1+{1\over x_1-\xi}{1\over y_1-\eta}\right)\left(1+{1\over x_2-\xi}{1\over y_2-\eta}\right)  &
{1\over x_1-\xi}{1\over y_1-\eta}{1\over x_2-\xi}{1\over y_2-\eta}  \cr
{1\over x_1-\xi}{1\over y_2-\eta}{1\over x_2-\xi}{1\over y_1-\eta} &
\left(1+{1\over x_1-\xi}{1\over y_2-\eta}\right)\left(1+{1\over x_2-\xi}{1\over y_1-\eta}\right)  \cr
}
\eeq
i.e.
\bea
 {\cal M}^{(2)}(x_1,x_2,y_1,y_2,\xi,\eta)
&=&\left(1+{1\over 2}\left({1\over x_1-\xi}+{1\over x_2-\xi}\right)\left({1\over y_1-\eta}+{1\over y_2-\eta}\right)
\right)\pmatrix{1&0\cr 0&1} \cr
&& +{1\over (x_1-\xi)(x_2-\xi)(y_1-\eta)(y_2-\eta)}\pmatrix{1+S&1\cr 1&1-S} \cr
\eea
where
\beq
S={1\over 2}(x_1-x_2)(y_1-y_2)
\eeq
Define the following orthogonal matrix (${\cal U}^{(2)}(x_1,x_2,y_1,y_2)\,{\cal U}^{(2)}(x_1,x_2,y_1,y_2)^t=\1$):
\beq
{\cal U}^{(2)}(x_1,x_2,y_1,y_2):={1\over \sqrt{2\l(\l-S)}}\,\pmatrix{1 & \l-S \cr S-\l & 1}
\virg
{\rm where}\,\,\l=\sqrt{1+S^2}
\eeq
one has:
\bea
{\cal M}^{(2)}(x_1,x_2,y_1,y_2,\xi,\eta)=
{\cal U}^{(2)}(x_1,x_2,y_1,y_2)\,
\L^{(2)}(x_1,x_2,y_1,y_2,\xi,\eta)\,
{\cal U}^{(2)}(x_1,x_2,y_1,y_2)^t
\eea
where $\L^{(2)}(x_1,x_2,y_1,y_2,\xi,\eta)=\diag(\L_+(x_1,x_2,y_1,y_2,\xi,\eta),\L_-(x_1,x_2,y_1,y_2,\xi,\eta)\,)$ with
\bea
\L_{\pm}(x_1,x_2,y_1,y_2,\xi,\eta)
&=& 1+{1\over 2}\left({1\over x_1-\xi}+{1\over x_2-\xi}\right)\left({1\over y_1-\eta}+{1\over y_2-\eta}\right) \cr
&& +{1\pm\l\over (x_1-\xi)(x_2-\xi)(y_1-\eta)(y_2-\eta)}
\eea
Eventualy, one gets:
\bea
&&{\int_{T(n)} dT\,\ee{-\Tr T T^\dagger}\, \Tr {1\over x_1-(X+T)}{1\over y_1-(Y+T^\dagger)}{1\over x_2-(X+T)}{1\over y_2-(Y+T^\dagger)}
\over \int_{T(n)} dT\, \ee{-\Tr T T^\dagger}}  \cr
&=&{1\over 2\l}\left(\prod_{i=1}^n \L_-(x_1,x_2,y_1,y_2,X_i,Y_i)-\prod_{i=1}^n \L_+(x_1,x_2,y_1,y_2,X_i,Y_i)\right)
\eea

\bea
&&{\int_{T(n)} dT\,\ee{-\Tr T T^\dagger}\, \left(1+\Tr {1\over x_1-(X+T)}{1\over y_1-(Y+T^\dagger)}\right)\left(1+\Tr{1\over x_2-(X+T)}{1\over y_2-(Y+T^\dagger)}\right)
\over \int_{T(n)} dT\, \ee{-\Tr T T^\dagger}}  \cr
&=&{1\over 2\l}\left((\l+S)\prod_{i=1}^n \L_+(x_1,x_2,y_1,y_2,X_i,Y_i)+(\l-S)\prod_{i=1}^n \L_-(x_1,x_2,y_1,y_2,X_i,Y_i)\right)
\eea

\section{Mixed correlation functions and biorthogonal polynomials}

Let us consider two polynomial potentials $V_1(x)$ and $V_2(y)$ .
Our goal is to compute the following matrix expectation values:
\beq
{\int_{H_n\times H_n}\, dM_1\, dM_2\, F_{\pi,\pi'}(\vec{x},\vec{y},M_1,M_2)\, \ee{-\Tr(V_1(M_1)+V_2(M_2)+ M_1 M_2)}
\over \int_{H_n\times H_n}\, dM_1\, dM_2\, \ee{-\Tr(V_1(M_1)+V_2(M_2)+ M_1 M_2)}}
\eeq

\subsection{Biorthonormal polynomials}

We recall here a few elementary notions about biorthogonal polynomials. More detailed descriptions can be found in particular in \cite{Mehta, mehta, BEmixed, BEHduality, BEHRH, BEformulaD}.

We introduce two families of polynomials $p_n(x)={1\over \sqrt{h_n}}x^n+O(x^{n-1})$, $q_n(y)={1\over \sqrt{h_n}}y^n+O(y^{n-1})$,
with the same leading coefficient ${1\over \sqrt{h_n}}$, and orthonormal with respect to the pairing:
\beq
(p_n,q_m) = \int\int dx\, dy\, p_n(x)\, q_m(y)\, \ee{-(V_1(x)+V_2(y)+xy)} = \delta_{nm}
\eeq
The integration path is a priori $\R\times \R$, but this condition can be relaxed (see \cite{Berto, BEHRH}).
When they exist, the biorthonormal polynomials are uniquely determined.

Since the biorthonormal polynomials form a basis, one can decompose $xp_n(x)$ onto the basis of $p_m(x)$ with $m\leq n+1$:
\beq
x\, p_n(x) = \sum_{m=0}^{n+1} Q_{nm}\, p_m(x)
\eeq
and similarly:
\beq
y\, q_n(y) = \sum_{m=0}^{n+1} P_{nm}\, q_m(y)
\eeq
$Q$ and $P$ are infinite matrices.
In the case where $V_2$ (resp. $V_1$) is a polynomial, then $Q$ (resp. $P$) is  a finite band matrix.

We also introduce the following $\infty\times n$ rectangular matrix:
\beq
\Pi_{n-1}:=\pmatrix{1 & &  \cr & \ddots  & \cr & & 1 \cr & &  \cr & 0 & \cr & &  }
\eeq
which is the projector onto the $n$ first polynomials.

\subsection{Mixed correlation functions}
\label{mixedcor}

\bt
\label{thmmixedcor}
\bea
&& {\int_{H_n\times H_n}\, dM_1\, dM_2\, F_{\pi,\pi'}(\vec{x},\vec{y},M_1,M_2)\, \ee{-\Tr(V_1(M_1)+V_2(M_2)+ M_1 M_2)}
\over \int_{H_n\times H_n}\, dM_1\, dM_2\, \ee{-\Tr(V_1(M_1)+V_2(M_2)+ M_1 M_2)}} \cr
&=& \sum_\rho {\cal U}_{\pi,\rho}(\vec{x},\vec{y})\,{\cal U}_{\pi',\rho}(\vec{x},\vec{y})
\det\left(\Pi_{n-1}^t :\L_\rho(\vec{x},\vec{y},Q,P^t):\Pi_{n-1}\right)
\eea
where for any function of two variables $f(\xi,\eta)$, we define $:f(Q,P^t):$ by putting the $Q$'s on the right of the $P$'s.
This is always possible in this case because $\L_\rho(\vec{x},\vec{y},\xi,\eta)$ is a rational function of $\xi$ and $\eta$.
\et

\proof{
It works as usual (see \cite{mehta, Mehta}), by writings Vandermonde determinants as:
\beq
\Delta(X) = \det(X_i^{j-1}) = \det(\sqrt{h_{j-1}}\,p_{j-1}(X_i))= \prod_{i=0}^{n-1} \sqrt{h_i}\,\sum_{\sigma} (-1)^\sigma \prod_i p_{\sigma(i)}(X_i)
\eeq
\beq
\Delta(Y) = \det(Y_i^{j-1}) = \det(\sqrt{h_{j-1}}\,q_{j-1}(Y_i))= \prod_{i=0}^{n-1} \sqrt{h_i}\,\sum_{\tau} (-1)^\tau \prod_i q_{\tau(i)}(Y_i)
\eeq
Then, we use eq.\ref{UnintLambdaperm}, i.e.
\bea
&& {1\over c_n\,{\td J}_n^2\,\pi^{n(n-1)\over 2}}\,\int_{H_n\times H_n}\, dM_1\, dM_2\, F_{\pi,\pi'}(\vec{x},\vec{y},M_1,M_2)\, \ee{-\Tr(V_1(M_1)+V_2(M_2)+ M_1 M_2)} \cr
&=& {1\over n!^2}\,\prod_{i=0}^{n-1} {h_i}\,\sum_{\rho\in \Sym(R)} {\cal U}_{\pi,\rho}(\vec{x},\vec{y})\,{\cal U}_{\pi',\rho}(\vec{x},\vec{y}) \sum_{\sigma,\tau,\nu\in \Sym(n)} (-1)^{\sigma\tau\nu}  \cr
&& \int \prod_i \L_\rho(\vec{x},\vec{y},X_i, Y_{\nu(i)}) p_{\sigma(i)}(X_i) \ee{-V_1(X_i)} q_{\tau\nu(i)}(Y_{\nu(i)}) \ee{-V_2(Y_{\nu(i)})}\ee{-X_i Y_{\nu(i)}} dX_i dY_{\nu(i)}\cr
&=& {1\over n!^2}\,\prod_{i=0}^{n-1} {h_i}\,\sum_{\rho\in \Sym(R)} {\cal U}_{\pi,\rho}(\vec{x},\vec{y})\,{\cal U}_{\pi',\rho}(\vec{x},\vec{y}) \sum_{\sigma,\tau,\nu\in \Sym(n)} (-1)^{\sigma\tau\nu}  \prod_{i=1}^n :\L_\rho(\vec{x},\vec{y},Q,P^t):_{\sigma(i),\tau\nu(i)}\cr
&=& \prod_{i=0}^{n-1} {h_i}\,\sum_{\rho\in \Sym(R)} {\cal U}_{\pi,\rho}(\vec{x},\vec{y})\,{\cal U}_{\pi',\rho}(\vec{x},\vec{y}) \sum_{\sigma\in \Sym(n)} (-1)^{\sigma}  \prod_{i=1}^n :\L_\rho(\vec{x},\vec{y},Q,P^t):_{i,\sigma(i)}\cr
&=& \prod_{i=0}^{n-1} {h_i}\,\sum_{\rho\in \Sym(R)} {\cal U}_{\pi,\rho}(\vec{x},\vec{y})\,{\cal U}_{\pi',\rho}(\vec{x},\vec{y})
\det{\left(\Pi_{n-1}^t:\L_\rho(\vec{x},\vec{y},Q,P^t):\Pi_{n-1}\right)}\cr
\eea

}

or, using the matricial determinant defined in def.\ref{defMdet}:
\bea
&& {\int_{H_n\times H_n}\, dM_1\, dM_2\, F_{\pi,\pi'}(\vec{x},\vec{y},M_1,M_2)\, \ee{-\Tr(V_1(M_1)+V_2(M_2)+ M_1 M_2)}
\over \int_{H_n\times H_n}\, dM_1\, dM_2\, \ee{-\Tr(V_1(M_1)+V_2(M_2)+ M_1 M_2)}} \cr
&=& \left( \Mdet \left(\Pi^t_{n-1}:{\cal M}^{(R)}(\vec{x},\vec{y},Q,P^t):\Pi_{n-1}\right)\right)_{\pi,\pi'}
\eea

Example: with $R=1$, we find:
\bea
&& {\int_{H_n\times H_n}\, dM_1\, dM_2\, (1+\Tr{1\over x-M_1}{1\over y-M_2})\, \ee{-\Tr(V_1(M_1)+V_2(M_2)+ M_1 M_2)}
\over \int_{H_n\times H_n}\, dM_1\, dM_2\, \ee{-\Tr(V_1(M_1)+V_2(M_2)+ M_1 M_2)}} \cr
&=& \det \left(\Pi^t_{n-1}\left(1+{1\over y-P^t}{1\over x-Q}\right)\Pi_{n-1}\right)
\eea
which is identical to what was found in \cite{BEmixed}.

\section{Conclusions}

In this article, we have shown that the hermitean 2-matrix model and the complex matrix model have the same loop equations.
In the gaussian case, that implies they are identical.
In case the weight is non--gaussian, the loop equations, which are recursion equations, determine all correlation functions when some initial conditions (moduli) are fixed.
The generalization of the hermitean 2-matrix model to homology classes of contours (as in \cite{BEHRH}), allows to have any arbitrary initial condidtions, so, there exists
a choice of homology class of contours for each set of initial conditions, i.e. for which the complex matrix model is  identical to the 2-hermitean matrix model.
Conversely, the initial conditions for the complex matrix model are not fully understood yet, they depend on how the complex matrix model is defined.
If the complex matrix model is only a formal integral defined by its large $n$ properties as in \cite{WiegZab, Zab}, initial conditions are associated to filling fractions, and can thus be chosen arbitrarily.
If the complex matrix model is defined as the result of a convergent integral for all $n$, it is not known yet how to find which homology class of contours it corresponds to.

\smallskip

The consequence of that identification, through diagonalization of hermitean matrices and Jordanization of complex matrices, yields an identity between
unitary group integrals and triangular matrices integrals, which seems to be a special case of the identification of $GL_n(\C)/T(n)$ and the quotient of $SU(n)$ by its Cartan subalgebra.
The nature of that identification needs to be further understood, in particular in terms of characters of both groups, and in terms of group representation theory, in terms of Weyl's character formula, or Harish-Chandra formulae.

\smallskip

The gaussian triangular matrix integrals are easily computed, and we thus get very explicit expressions for all expectation values of the type which were studied by
Shatashvili \cite{shata}.
In particular, we have provided a new proof of the Itzykson-Zuber-Harish-Chandra integral, as well as Morozov's integral.
The key piece in this computation is that the matrices ${\cal M}$ commute together. This fact seems to be related to some Yang-Baxter relations, and it would be interesting to understand how.

\smallskip

It would be interesting also to understand these formulae in the framework of Duistermaat-Heckman semiclassical theories \cite{DH}.

\smallskip

Then, we have been able to perform the integral over eigenvalues, in a way very similar to what was done in \cite{BEmixed}, i.e.
in terms of $n\times n$ determinants.
It would then be interesting to rewrite these $n\times n$ determinants in terms of determinants of size independent of $n$, using kernels,
as it is known for non-mixed expectations values (see \cite{bergere, bergereLH, eynardmehta}).

\bigskip
{\noindent \bf Aknowledgements:}
The authors want to thank  F. David, P. Di Francesco, J.B. Zuber for stimulating discussions.
One of the authors (B.E.) wants to thank the european network Enigma (MRTN-CT-2004-5652).
A. P-F. wants to thank the SPhT Saclay for its hospitality when part of this work was being conducted,
and the support of CIRIT grant 2001FI-00387.

\setcounter{section}0
\setcounter{equation}0
\appendix{Gaussian integrals}
\label{apintRintCpol}

Let $\delta=\alpha_1\alpha_2-\gamma^2$.

\medskip

{\bf\noindent Real integrals:}
\beq
\int_{\R\times \R}\,dx\,dy\,\ee{-({\alpha_1\over 2}x^2 + {\alpha_2\over 2}y^2 + \gamma xy)}
= {2\pi\over \sqrt{\delta}}
\eeq

\bea\label{intRRxkyl}
{\int_{\R\times \R}\,dx\,dy\,x^k y^l\,\ee{-({\alpha_1\over 2}x^2 + {\alpha_2\over 2}y^2 + \gamma xy)}
\over \int_{\R\times \R}\,dx\,dy\,\ee{-({\alpha_1\over 2}x^2 + {\alpha_2\over 2}y^2 + \gamma xy)}}
&=& 0 \qquad {\rm if}\,\, k+l \,\,{\rm is\,\, odd} \cr
&=&\left({\sqrt\delta\over 2\pi}\right)\,\left(-2{\d\over \d\alpha_1}\right)^{{k-l\over 2}}\left(-{\d\over \d\gamma}\right)^{l}\,{\left({2\pi\over \sqrt\delta}\right)} \qquad {\rm if}\,\, k\geq l \cr
&=&\left({\sqrt\delta\over 2\pi}\right)\,\left(-2{\d\over \d\alpha_2}\right)^{{l-k\over 2}}\left(-{\d\over \d\gamma}\right)^{k}\,{\left({2\pi\over \sqrt\delta}\right)} \qquad {\rm if}\,\, k\leq l \cr
\eea

\beq
\int_{D_n(\R)\times D_n(\R)}\,dX\,dY\,\ee{-\Tr({\alpha_1\over 2}X^2 + {\alpha_2\over 2}Y^2 + \gamma XY)}
= \left({2\pi\over \sqrt{\delta}}\right)^{n}
\eeq

\medskip

{\bf\noindent Complex integrals:}

\beq
\int_{\C}\,dx\,\ee{-({\alpha_1\over 2}x^2 + {\alpha_2\over 2}\ovl{x}^2 + \gamma x\ovl{x})}
= {\pi\over \sqrt{-\delta}}
\eeq

\bea\label{intCCxkyl}
{\int_{\C}\,dx\,\,x^k\, \ovl{x}^l\,\ee{-({\alpha_1\over 2}x^2 + {\alpha_2\over 2}\ovl{x}^2 + \gamma x\ovl{x})}
\over \int_{\C}\,dx\,\ee{-({\alpha_1\over 2}x^2 + {\alpha_2\over 2}\ovl{x}^2 + \gamma x\ovl{x})}}
&=& 0 \qquad {\rm if}\,\, k+l \,\,{\rm is\,\, odd} \cr
&=&\left({\sqrt{-\delta}\over \pi}\right)\,\left(-2{\d\over \d\alpha_1}\right)^{{k-l\over 2}}\left(-{\d\over \d\gamma}\right)^{l}\,{\left({\pi\over \sqrt{-\delta}}\right)} \qquad {\rm if}\,\, k\geq l \cr
&=&\left({\sqrt{-\delta}\over \pi}\right)\,\left(-2{\d\over \d\alpha_2}\right)^{{l-k\over 2}}\left(-{\d\over \d\gamma}\right)^{k}\,{\left({\pi\over \sqrt{-\delta}}\right)} \qquad {\rm if}\,\, k\leq l \cr
\eea

\beq
\int_{D_n(\C)}\,dX\,\ee{-\Tr({\alpha_1\over 2}X^2 + {\alpha_2\over 2}\ovl{X}^2 + \gamma X\ovl{X})}
= \left({\pi\over \sqrt{-\delta}}\right)^{n}
\eeq

\medskip

\bl\label{lemmaappendix}
Let $\om(X,Y)$, be a polynomial in all its variables $X_1,\dots,X_n$ and  $Y_1,\dots,Y_n$, one has:
\beq
{\int_{D_n(\R)\times D_n(\R)}\,dX\,dY\,\om(X,Y)\,\ee{-\Tr({\alpha_1\over 2}X^2 + {\alpha_2\over 2}Y^2 + \gamma XY)}
\over \int_{D_n(\R)\times D_n(\R)}\,dX\,dY\,\ee{-\Tr({\alpha_1\over 2}X^2 + {\alpha_2\over 2}Y^2 + \gamma XY)}}
= {\int_{D_n(\C)}\,dX\,\om(X,\ovl{X}) \, \,\ee{-\Tr({\alpha_1\over 2}X^2 + {\alpha_2\over 2}\ovl{X}^2 + \gamma X\ovl{X})}
\over \int_{D_n(\C)}\,dX\, \,\ee{-\Tr({\alpha_1\over 2}X^2 + {\alpha_2\over 2}\ovl{X}^2 + \gamma X\ovl{X})}}
\eeq
\el

\proof{
Eqs \ref{intRRxkyl} and \ref{intCCxkyl} show that it is true for $n=1$. By decomposing $\om$ into monomials, the integral decouples into a product of $n=1$ type integrals.
}

\appendix{Some Commutations}

\bt\label{Thmcommutbis}
The matrix ${\cal M}^{(R)}(\vec{x},\vec{y},\xi,\eta)$ commutes with the matrix ${\cal A}(\vec{x},\vec{y})$ defined by:
\beq
\left\{
\begin{array}{l}
{\cal A}_{\pi,\pi}(\vec{x},\vec{y}) := \sum_i x_i y_{\pi(i)} \cr
{\cal A}_{\pi,\pi'}(\vec{x},\vec{y}) := 1 \quad {\rm if}\,\, \pi\pi'^{-1}={\rm transposition}  \cr
{\cal A}_{\pi,\pi'}(\vec{x},\vec{y}) := 0 \quad {\rm otherwise}
\end{array}
\right.
\eeq
\et

\bt\label{Thmcommutter}
The matrices ${\cal A}^{\alpha,\beta}(\vec{x},\vec{y})$ defined by:
\bea
{\cal A}^{\alpha,\beta}_{\pi,\pi'}(\vec{x},\vec{y})
&:=& \delta_{\beta,\pi(\alpha)} \prod_{i\neq \alpha} \left(\delta_{\pi(i),\pi'(i)}+{1\over x_\alpha-x_i}{1\over y_\beta-y_{\pi(i)}}\right) \cr
&& + {1-\delta_{\beta,\pi(\alpha)}\over (x_\alpha-x_{\pi^{-1}(\beta)})(y_\beta-y_{\pi(\alpha)})}\, \prod_{i\neq \alpha,\pi^{-1}(\beta)} \left(\delta_{\pi(i),\pi'(i)}+{1\over x_\alpha-x_i}{1\over y_\beta-y_{\pi(i)}}\right) \cr
\eea
commute together for all $\alpha,\beta$. They also commute with ${\cal M}(\vec{x},\vec{y},\xi,\eta)$ and with ${\cal A}(\vec{x},\vec{y})$.
\et
One has:
\beq
{\cal M}^{(R)}(\vec{x},\vec{y},\xi,\eta) = \1+\sum_{\alpha,\beta} {1\over (\xi-x_\alpha)(\eta-y_\beta)}\,{\cal A}^{\alpha,\beta}(\vec{x},\vec{y})
\eeq

\end{document}